  \documentclass[prl,aps,twocolumn,showpacs]{revtex4}
\usepackage[dvips]{graphicx}
\usepackage{dcolumn}%
\usepackage{amsmath}%
\usepackage{amsfonts}%
\usepackage{amssymb}
 \usepackage[usenames]{color}
\providecommand{\U}[1]{\protect\rule{.1in}{.1in}}
\newcommand{\be}{\begin{equation}}
\newcommand{\ee}{\end{equation}}
\newcommand{\bea}{\begin{eqnarray}}
\newcommand{\eea}{\end{eqnarray}}
\newcommand{\bt} {\begin{tabular}}
\newcommand{\et} {\end{tabular}}
\newcommand{\nn}{ \nonumber}

\newcommand{\ba} {\begin{array}}
\newcommand{\ea} {\end{array}}
\begin{document}

\title{Length-dependent Seebeck  effect in single-molecule junctions  beyond linear response regime}

\author{  Natalya A. Zimbovskaya}

\affiliation
{Department of Physics and Electronics, University of Puerto 
Rico,  Humacao, Puerto Rico 00791, USA}

\begin{abstract}
In the present work we theoretically study characteristics  of nonlinear Seebeck effect  in  a single-molecule junction with chain-like bridge of an arbitrary length. We have employed tight-binding models to compute electron transmission trough the system. We concentrate on analysis of dependences of thermovoltage $V_{th} $ and differential thermopower $ S $ on the bridge length. It is shown that $ V_{th} $ becomes stronger and $ S $ grows as the bridge lengthens.  We discuss the effects of the bridge coupling to the electrodes and of specific characteristics of terminal sites on the bridge on the length-dependent $ V_{th} $ and $ S $ which appear when the system operates beyond linear response regime.
		\end{abstract}

%% \pacs{72.15.Gd,71.18.+y}%%{71.18.+y, 71.20-b, 72.55+s}

\date{\today}
\maketitle

\section{I. Introduction}

Molecular junctions are well known species of tailored nanoscale systems which consist of a couple of metallic/semiconducting electrodes linked by a single  molecule (or by several molecules). Presently, transport properties of molecular junctions  are intensively studied due to their possible applications in molecular electronics \cite{1,2,3}. Thermoelectric properties of such systems are especially interesting for they hold promise for enhanced efficiency of heat-to-electric energy conversion \cite{4,5,6,7,8,9}. In this work, we concentrate on Seebeck effect in molecular junctions \cite{10}. As commonly known, a difference in the temperatures of electrodes $ \Delta \theta $  induces a charge current which flows through the system. Seebeck effect is generation of the voltage $ V_{th} $ which stops this current provided that $ \Delta \theta $ is kept constant. When $ \Delta \theta \ll \theta \ (\theta $ being the average temperature in the system) the system operates within the linear response regime, so $ V_{th} $   is proportional to  $\Delta \theta, $  coefficient of proportionality being the system thermopower.

Numerous works are focused on exploring properties of the thermopower of single-molecule junctions. It was shown that the thermopower may be affected by molecular 
vibrations \cite{11,12,13,14,15,16,17}, by effects of molecular bridge geometry \cite{18,19,20,21,22} by interactions between electrons participating in transport \cite{23,24,25,26,27,28,29,30} and by photons \cite{31}. Under certain conditions (e.g. in single-molecule junctions with ferromagnetic electrodes) spin polarization of electrons may significantly influence the thermopower  \cite{32,33,34,35,36}. Also, the thermopower may be affected due to dissipative interactions of traveling electrons with the ambience \cite{37,38,39,40,41,42} and due to effects of quantum interference \cite{43,44}. 

It was discovered that both thermopower and electron conductance of single-molecule junctions may strongly depend on the length of the molecular bridge linking the electrodes. Length-dependent conductance and thermopower are usually observed and studied in junctions whose molecular linkers are chain-like structures consisting of several identical units (e.g. benzene or phenyl rings) \cite{12,19,20,21,22,45,46,47,48,49,50,51,52,53}. In the most of experiments, the thermopower appears to be proportional to the molecular bridge length. However, in recent experiments performed on single-molecule junctions with oligophenyl and alcane chain-like linkers, more complex relationships between the thermopower and molecular bridge length were demonstrated \cite{50}.

As the temperature  difference between the electrodes $ \Delta \theta $ increases, the system may switch to a nonlinear regime of operation. Nonlinear Seebeck effect was already observed in semiconductor quantum dots and single-molecule junctions \cite{50,54}. In several recent works the discussion of Seebeck effect in these systems was extended beyond linear in $ \Delta \theta $  response \cite{10,12,32,55,56,57,58,59,60,61,62}. However, dependences  $ V_{th} $ on the molecular bridge length beyond linear response regime  were not thoroughly analyzed so far. The purpose of the present work is to contribute to studies of molecular thermoelectric transport through single-molecule junctions by analyzing length-dependences of thermovoltage $ V_{th} $ at sufficiently high values of $ \Delta \theta, $ outside the scope of applicability of the linear response theory.

\section{ii. model and main equations}

In the following analysis we are basing on a model where the molecular bridge is represented by a periodical chain including $ N $ sites. Each site is assigned an on-site energy $ E_i $ and coupled to  its nearest neighbors. We consider two versions of this model. Within the first version, we assume that all on-site energies are equal $ (E_i = E_0,\ 1 \leq i \leq N) $ and the coupling between adjacent sites has the same strength (characterized by the coupling parameter $\beta) $ over  the entire chain . This model could be applied to describe molecular bridges where $ \pi - \pi $ dominates electron transport. Then the parameter $ \beta $ characterizes the coupling between adjacent $ \pi $ orbitals \cite{63}.  Within the second version, we separate out two sites at the ends of the chain. These terminal sites are singled out because they may affect dependences of thermoelectric transport characteristics of single-molecule junctions on the molecular bridge length as was recently demonstrated \cite{50,53}. We set on terminal states on-site energies $ E_1 = E_N = \epsilon $ which differ from $ E_0. $ Also, we assume that these sites are coupled to their neighbors with the coupling strength $ \delta $ different from $ \beta . $ 
    Within this version of tight-binding model, the terminal sites are associated with gateway states representing bonds  between terminal atoms on the bridge and electrodes as it happens, for example, in junctions with oligophenyl and alcane bridges \cite{50}. Within both versions of the employed model, the ends of the chain are coupled to electrodes through imaginary self-energy terms $ -i\Gamma/2 $ which are supposed to be energy-independent. In further analysis  we assume that coherent electron tunneling is a predominant transport mechanism.
		
When electrodes are kept at different temperatures $(\theta_L $ and $ \theta_R ,$ respectively) electric current flows through an unbiased junction. For a symmetrically coupled junction considered in the present work, this current is given by the Landauer expression:
\be
I _{th} = \frac{e}{\pi\hbar} \int \tau (E, \theta_L,\theta_R) \big [f^L(E, \theta_L) - f^R (E,\theta_R) \big] dE   \label{1}
\ee 		
where $ f^{L,R} (E,\theta_{L,R}) $ are  Fermi distribution functions for the electrodes  and the electron transmission function $ \tau(E, \theta_L,\theta_R)$ may be presented in the form:
\be
\tau(E, \theta_L,\theta_R) = \frac{\Gamma^2}{4} \big|G_{1N} (E, \theta_L,\theta_R) \big|^2  \label{2} .
\ee
Here, $ G_{1N} $ is the corresponding matrix element of the retarded Green,s function for the bridge:
\be
G = (E - H - i\Gamma)^{-1}   \label{3}
\ee
Employing the first version of the accepted model, the Hamiltonian $ H $ is represented by $ N \times N $ matrix:
\be 
H =  \left [\ba{cccccc}
\tilde E_0 - \frac{i\Gamma}{2} & \beta & 0 & 0 & \dots & 0
 \\
\beta & \tilde E_0 & \beta & 0 & \dots & 0
\\
0  & \beta & \tilde E_0 & \beta & \dots & 0
\\
\dots & \dots & \dots & \dots & \dots & \dots
\\
0 & 0 & \dots & \beta & \tilde E_0 & \beta
\\
0 & 0 & \dots &  0 & \beta & \tilde E_0 - \frac{i\Gamma}{2}
\\ 
\ea \right ] .  \label{4}
\ee
	
Using Eqs. (\ref{3}), (\ref{4}), an expression for $ G_{1N} $ may be derived in the form \cite{63,64}:
\be
G_{1N} (E,\theta_L,\theta_R) = \frac{\beta^{N-1}}{\Delta_N(E,\Gamma)}  \label{5}
\ee
where the determinant $ \Delta_N(E,\Gamma) $ equals:
\begin{align}
&  \Delta_N(E,\Gamma) =  \frac{1}{2^{N + 1}\zeta}             \label{6}
\\  \times &
\Big[(\lambda + \zeta)^{N-1} (\lambda + \zeta + i\Gamma)^2 - 
(\lambda - \zeta)^{N-1} (\lambda - \zeta + i\Gamma)^2 \Big] \nn  .
\end{align}		
In these expressions, $ \lambda = E - \tilde E_0,\  \zeta = \sqrt{\lambda^2 - 4 \beta^2} $ and $ \tilde E_0 $ is the on-site energy renormalized in the presence of the temperature difference $ \Delta\theta = \theta_L - \theta_R $ \cite{60}. This renormalization makes $ \Delta_N (E,\Gamma) $ and, correspondingly, $ G_{1N} $ dependent on the electrodes temperatures.  When the on-site energy $ E_0 $ takes on values comparable with thermal energy $ k \Delta \theta \ (k $ being the Boltzmann's constant) the renormalization may lead to noticeable changes in the electron transmission. However, when $ E_0  \gg k \Delta \theta $ the effect of temperature on the electron transmission becomes negligible.

When the terminal "gateaway" sites different from the remaining sites in the chain are taken into consideration, the Hamiltonian matrix accepts the form similar to that given by Eq. (\ref{4})  where the Hamiltonian $ H $ is represented by $ N \times N $ matrix:
\be 
H =  \left [\ba{cccccc}
\tilde \epsilon - \frac{i\Gamma}{2} & \delta & 0 & 0 & \dots & 0
 \\
\delta & \tilde E_0 & \beta & 0 & \dots & 0
\\
0  & \beta & \tilde E_0 & \beta & \dots & 0
\\
\dots & \dots & \dots & \dots & \dots & \dots 
\\
0 & 0 & \dots & \beta & \tilde E_0 & \delta
\\
0 & 0 & \dots & \dots & \delta & \tilde \epsilon - \frac{i\Gamma}{2}
\\ 
\ea \right ].  \label{7}
\ee
In this expression, $ \tilde \epsilon $ is also renormalized due to the presence of $ \Delta \theta, $ and $ G_{1N} $ equals $(N \geq 3): $
\be
G_{1N} = \frac{\delta^2 \beta^{N -3}}{\tilde \Delta_N(E, \Gamma)}.  \label{8}
\ee
The determinant $ \tilde \Delta_N (E,\Gamma) $ is given by the expression \cite{53}:
\begin{align}
 & \tilde \Delta_N (E,\Gamma)  
\nn\\ = &  
\Delta_N (E,\Gamma) + (\alpha - \lambda)(\alpha + \lambda + i \Gamma)
 \Delta_{N-2} (E,0)
\nn \\    & + 
\big[(\beta^2 - \delta^2)(\alpha + \lambda + i\Gamma) - (\alpha - \lambda)(\beta^2 + \delta^2)\big] 
\nn\\ & \times
\Delta_{N-3}(E,0)  - (\beta^4 - \delta^4) \Delta_{N-4} (E,0)   \label{9}
\end{align}
where $ \alpha = \tilde E_0 - \tilde \epsilon, \ \Delta_N(E,\Gamma) $ is described by Eq. (\ref{6}), and other determinants are obtained from Eq. (\ref{6}) by setting $ \Gamma = 0. $ In particular, $\Delta_1(E,0) = 1$ and $\Delta_{-1}(E,0) = 0.$ 

As follows from  Eq. (\ref{1}) for the thermocurrent in an unbiased junction, only those charge carriers whose energies are very close to the chemical potential of electrodes $ \mu $ can contribute to $ I_{th}. $  Therefore, the specifics of electron transmission profile near $ E = \mu $ may considerably influence characteristics of thermoelectric transport. In Fig. 1 we display the electron transmission behavior for a simple chain of identical sites and for a chain including terminal sites with different properties. In plotting the curves presented in the left panel we accepted  the same values for relevant parameters as those derived for a single-molecule junction with an oligophenyl bridge in the previous work \cite{50}, namely: $ E_0 = -4.47 eV, \ \epsilon = - 1.85 eV,\  \Gamma = 2.85 eV,\ \delta = 2.28 eV $ and $ \beta = 1.27 eV. $ In plotting electron transmission profiles for a simple chain of  identical sites shown in the right panel we assumed that $ E_0 = -4.5 eV $ and $ \beta = 2.2 eV. $ In both cases the relevant energies take on much greater values than the thermal energy $ k\Delta \theta $ provided that the temperatures $ \theta_L $ and  $ \theta_R $ are of the same order or below the room temperature. This gives grounds to disregard the renormalization of on-site energies occurring due to the thermal gradient applied across the system. Consequently,  the electron transmission becomes temperature independent. We observe that gateaway states associated with the terminal sites on the bridge affect electron transmission distorting its symmetrical profile typical for a simple chain. Assuming that $ \mu = 0, $ the presence of these states mostly affects HOMO which serves as the channel for thermally induced transport. Peaks associated with HOMO become significantly broader and smoother than other resonance features appearing in both panels of Fig. 1. The distorted HOMO may bring noticeable changes in the behavior of length-dependent thermopower within the linear response regime as was observed in experiments \cite{50}.

\begin{figure}[t] %%% fig. 1
\begin{center}
 \includegraphics[width=8.9cm,height=4.3cm]{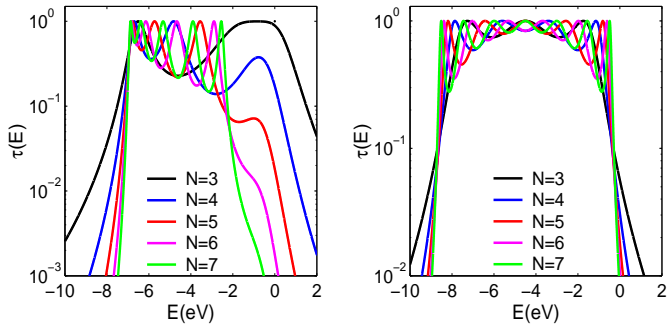}
\caption{Electron transmission profiles affected due to the presence of gateaway states on the molecular bridge (left panel) and electron transmission profiles for the bridge simulated by a periodical chain of identical sites (right panel). The curves are plotted assuming that $ E_0 = - 4.47 eV,\ \epsilon = - 1.85 eV,\ \Gamma = 2.86 eV,\  \delta = 2.28 eV,\  \beta = 1.27 eV $ (left panel) and $ E_0 = - 4.5 eV,\  \Gamma = 2.86 eV,\   \beta = 2.2 eV $ (right panel) at $ k \theta = 0.016 eV. $
}
 \label{rateI}
\end{center}\end{figure}

\section{iii. Results and Discussion}

In this Section, we present some numerical results derived by employing the outlined formalism. These results were obtained assuming that the right electrode is cooler than the left one $(\theta_L > \theta_R), $ and it is kept at a constant temperature whereas the temperature of the left electrode varies. The thermoelectric voltage appearing in an unbiased system may be computed by solving the equation:
\be
\int \tau(E) \big[f^L(E,V_{th},\theta_L) - f^R(E,V_{th},\theta_R) \big] = 0.  \label{10}
\ee
In this equation, we omit the electron transmission dependences on both the temperature difference $ \Delta \theta $ and the voltage $ V_{th}. $ . As discussed above, this approximation is a sound one for sufficiently strongly coupled molecular junctions where all relevant energies including the coupling parameter $ \Gamma $ exceed both $ k \Delta \theta $ and $ V_{th} .$ 

It is known that the thermovoltage takes on positive/negative values depending on the nature of charge carriers (electrons/holes) involved in the transport via corresponding transport channels. By assuming that $ \mu_L  = \mu_R = 0 $ and using the same values of all relevant energies which were used to compute electron transmission functions, we predetermine HOMO to be transport channel and $ V_{th} $ to take on negative values at sufficiently small $ \Delta \theta. $ As shown in Fig. 2, the increase in $ \Delta \theta $ leads to $ V_{th} $ decrease. At certain value of $ \Delta \theta $ the thermovoltage reaches its minimum. Further increase of the temperature difference first reduces  $ V_{th} $ magnitude and then brings the reversal of its polarity. Temperature dependences of the differential thermopower $ S = - \partial V_{th}/\partial \theta $ displayed in the right panel of Fig. 2 agree with those of $ V_{th}. $

\begin{figure}[t] %%% fig. 2
\begin{center}
\includegraphics[width=8.9cm,height=4.5cm]{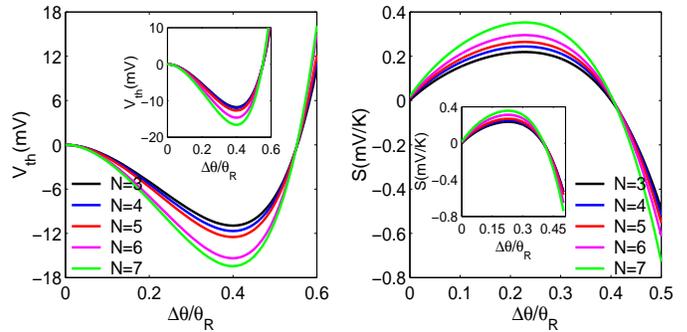}
\caption{Thermovoltage (left panel)  and differential thermopower (right panel) of a single-molecule junction as functions of temperature. The curves are plotted for different numbers of sites on the chain representing the moleculr bridge for  $ k\theta_R = 0.016eV,\  \mu = 0,\  E_0 = - 4.47eV, \  \epsilon = -1.85 eV,\  \Gamma = 2.86 eV,\  \delta = 2.28 eV,\  \beta = 1.27 eV. $ Insets show temperature dependences of $ V_{th} $ and $ S $ computed for the junction where the bridge is represented by a periodic chain of identical sites. The curves displayed in the insets are plotted assuming $k\theta_R = 0.016 eV,\  \mu = 0,\  E_0 = -4.5 eV,\  \Gamma = 2.86 eV,\  \beta = 2.2eV.$
}
 \label{rateI}
\end{center}\end{figure}

The described $ V_{th} $ behavior is similar to that reported for single-level quantum dots \cite{59,60} and may be explained as follows. The difference in the electrodes temperatures drives electrons from the hot (left) electrode to the cool one (right) thus creating a thermally-induced current. To suppress this current, a negative thermovoltage emerges whose magnitude increases as $ \Delta \theta $ grows. At the same time, the step in the Fermi distribution function for the hot (left) electrode is being partially smoothed out when  $ \theta_L $ rises. It opens the way for holes which start to flow towards the cool electrode. At a certain value of $ \Delta \theta $ the hole flux completely counterbalances the electron flux, and $ V_{th} $ becomes zero. When $ \Delta \theta $ rises above this value, the hole flux predominates over the electron flux so $ V_{th} $ takes on positive values indicating the reversal of thermovoltage polarity.

The results presented in Figs. 2,3 show that $ V_{th} $ and $ S $ depend on the molecular bridge length. Since a single molecular orbital  (HOMO in the considered case) participates in transport, the thermovoltage becomes length-dependent due to the effect of the bridge length on the HOMO profile in the vicinity of the electrodes chemical potential. This effect may be rather significant as shown in the Fig. 1. One observes that the transmission peaks become sharper and narrower as the number of sites in the chain increases. The same is valid for the spectral function directly representing molecular orbitals. The sharper HOMO profile becomes the stronger charge curriers fluxes via HOMO emerge, and the higher voltage is required to stop them. It is demonstrated in Fig. 2,3. One observes that at each fixed value of $ \Delta \theta, $ the thermovoltage magnitude accepts higher values for longer molecular chains.

\begin{figure}[t] %%% fig. 3
\begin{center}
\includegraphics[width=8.9cm,height=4.3cm]{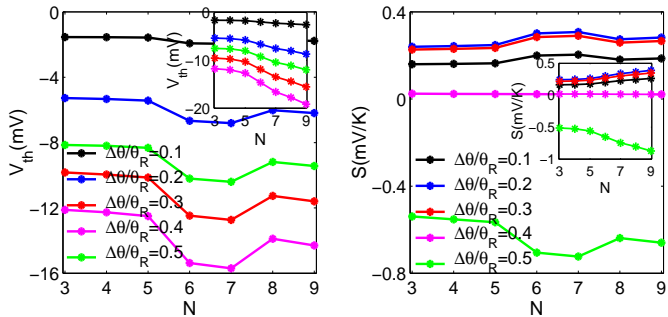}
\caption{Thermovoltage (left panel)  and differential thermopower (right panel) of a single-molecule junction as functions of the bridge length. The curves are plotted assuming that $ k\theta_R = 0.016eV,\  \mu = 0,\  E_0 = -4.47 eV,\  \epsilon = - 1.85 eV,\  \Gamma = 2.86 eV,\  \delta = 2.28 eV, \  \beta = 1.27 eV. $ Insets show length-dependent thermovoltage and differential thermopower computed for a junction with the  bridge represented by the chain of identical sites. The displayed curves are plotted assuming $ k\theta_R = 0.016 eV,\  \mu = 0,\  E_0 =- 4.5 eV,\  \Gamma = 2.86 eV,\  \beta = 2.2 eV. $
}
 \label{rateI}
\end{center}\end{figure}

Gateaway states may significantly affect temperature dependences of $ V_{th} $ and $ S $  for relatively short bridges.. Specifically, as shown in Fig. 2, a considerable separation occurs between $ V_{th} $ versus $ \theta $ curves representing short chains $ (3 \leq N \leq 5) $ and those corresponding to longer ones $( N > 5) .$ Also, dips appear on $ V_{th} $ versus $ N $ curves distorting general trends. These dips  indicate that a certain length of the molecular bridge may allow for the most favorable conditions for heat-to-electric energy conversion. For the chosen values of relevant parameters, the strongest $ V_{th} $ is reached at $ N = 7 $ for all values of $ \Delta \theta. $ At the same time, in the case of a simple chain of identical sites, the curves representing temperature dependeces of the thermovoltage plotted for different numbers of sites are "regularly" arranged so the distances between adjacent curves are nearly equal. Also, at each fixed value of $ \Delta \theta, \  V_{th}$  becomes stronger as the number of sites in a simple chain increases. We remark that the effect of gateaway states pronounced for short bridges quickly fades away as the chain becomes sufficiently long.

Another factor which can influence HOMO profile and, consequently, the thermovoltage is the coupling between the molecular bridge and elecrodes. We have analyzed the effect of coupling strength $ \Gamma $ on both the thermovoltage and the differential thermopower assuming that the bridge is simulated by a chain of identical sites. Obtained results shown in Fig. 4 give grounds to hypothesize that some values of coupling strength provide  better conditions for heat-to-electricity conversion than others. One observes that for strongly coupled junctions as well as for weakly coupled ones, $ V_{th} $ and $ S $ take on rather close values  which slowly vary as the number of sites increases. However, an intermediate coupling strength $(\Gamma = 0.8eV) $ brings a well pronounced rise in $ V_{th} $ and $ S $ when the bridge is not too short $( N > 5). $ In practical single-molecule junctions the coupling of electrodes to the bridge strongly depends on the contact chemistry \cite{19,20}.  For example, rotation of aromatic end groups in CSW-470-bipyridine molecular linkers was shown to significantly affect the coupling strength $ \Gamma $ in the corresponding molecular junctions \cite{19}.

\begin{figure}[t] %%% fig. 4
\begin{center}
\includegraphics[width=8.9cm,height=4.3cm]{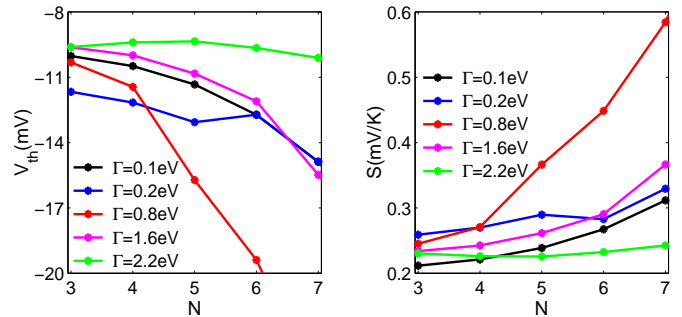}
\caption{Length-dependent thermovoltage (left panel) and differential thermopower (right panel) plotted at different values of the coupling strength $ \Gamma $ assuming that the molecular bridge in a single-molecule junction is simulated by a periodical chain of identical sites. In plotting the curves we set $ k\theta_R = 0.016 eV,\  \Delta\theta/\theta_R = 0.3,\  \mu = 0,\  E_0 = -4.5 eV,\  \beta = 2.2 eV.$  
}
 \label{rateI}
\end{center}\end{figure}

\section{iv. Conclusion}

In the present work we theoretically analyzed Seebeck effect in single-molecule junctions including long chain-like bridges. The analysis was carried out provided that the difference between the electrodes temperatures takes on considerable values, so the system operates beyond linear response regime. We focused on the effect of the molecular bridge length on the thermovoltage.

To compute the thermovoltage we have employed a tight-binding model simulating the bridge by a chain including an orbitrary number of  sites. Actually, we have used two versions of the model. Within the first version, all sites were supposed to be identical. Within the second version, we implied that terminal sites on the chain are characterized with different on-site energies and are coupled to their neighbors with a different strength. In choosing this model, we were inspired by  already reported results indicating that the presence of such sites with specific characteristics may significantly affect length dependences of thermopower in single-molecule junctions within the linear in $ \Delta \theta $ regime \cite{50}. 

Although simple, the adopted model allows to qualitatively study length  dependences of the thermovoltage and differential thermopower. Obtained results confirm that character of length dependences of $ V_{th} $ and$ S $ are determined by the profile of the peak in the electron transmission associated with the molecular orbital which serves as the transport channel for thermally induced electron transport. As the molecular bridge lenthens, $ V_{th} $ becomes stronger and $ S $ grows.

These general trends may be violated due to the influence of gateaway states which 
may considerably distort the electron transmission lineshapes. This results in minima/maxima on $ V_{th} $ versus $ N $ and $  S $ versus $ N $ curves. The influence of gateaway states may be considerable in the case of relatively short chains, but it fades away as the chain lengthens.  Also, the length-dependent thermovoltage may suffer significant changes when the coupling of the molecular bridge to the electrodes varies. It follows from the present analysis that at certain values of the coupling strength $ \Gamma $ the thermovoltage strengthens, especially when the bridge becomes sufficiently long, which agrees with the results on linear Seebeck effect in single-molecule junctions reported in several earlier works.

Finally, in this work we used simple tight-binding models to describe single-molecule junctions, and a simple computational method was employed to compute  relevant Green's functions. We realize limitations of the adopted models. Nevertheless, we believe that our results capture some  essential physics and may be useful for better understanding of Seebeck effect in single-molecule junctions beyond linear response regime.

\vspace{4mm}

 {\bf Acknowledgments:}
The author  thank  G. M. Zimbovsky for help with the manuscript preparation. This work was supported  by  NSF-DMR-PREM 1523463.

 \end{document}